\documentclass[
    showpacs,
    preprintnumbers,
    nofootinbib,
    prd,
    superscriptaddress,
    10pt,
    notitlepage,
    aps,
    twocolumn]{revtex4-1}
\usepackage{amssymb}
\usepackage{amsmath}
\usepackage{amsfonts}
\usepackage{amsbsy}

\usepackage{newtxtext}
\usepackage{newtxmath}

\usepackage{epsfig}
\usepackage{graphicx}
\usepackage{epstopdf}

\usepackage[linktocpage,breaklinks]{hyperref}
\usepackage[usenames,dvipsnames]{color}
\usepackage{empheq}
\hypersetup{colorlinks=true, citecolor=NavyBlue,
            linkcolor=NavyBlue, urlcolor=NavyBlue}
\usepackage{mathrsfs}
\usepackage{tensor}

\begin{document}

\title{On the stability of scalarized black hole solutions in scalar-Gauss-Bonnet gravity}

\author{Hector O. Silva}
\email{hector.okadadasilva@montana.edu}
\affiliation{eXtreme Gravity Institute,
Department of Physics, Montana State University, Bozeman, MT 59717 USA}

\author{Caio F.~B.~Macedo}
\email{caiomacedo@ufpa.br}
\affiliation{Campus Salin\'opolis, Universidade Federal do Par\'a,
Salin\'opolis, Par\'a, 68721-000, Brazil}

\author{Thomas P. Sotiriou}
\email{thomas.sotiriou@nottingham.ac.uk}
\affiliation{School of Mathematical Sciences,
University of Nottingham, University Park, Nottingham, NG7 2RD, UK}
\affiliation{School of Physics and Astronomy,
University of Nottingham, University Park, Nottingham, NG7 2RD, UK}

\author{Leonardo Gualtieri}
\email{leonardo.gualtieri@roma1.infn.it}
\affiliation{Dipartimento di Fisica ``Sapienza''
Universit\`a di Roma \& Sezione INFN Roma1,
Piazzale Aldo Moro 5, 00185, Roma, Italy}

\author{Jeremy Sakstein}
\email{sakstein@physics.upenn.edu}
\affiliation{Center for Particle Cosmology,
Department of Physics and Astronomy,
University of Pennsylvania,
209 S. 33rd St., Philadelphia, PA 19104, USA}

\author{Emanuele Berti}
\email{berti@jhu.edu}
\affiliation{Department of Physics and Astronomy,
Johns Hopkins University, 3400 N. Charles Street, Baltimore, MD 21218, USA}

\begin{abstract}
Scalar-tensor theories of gravity where a new scalar degree of
freedom couples to the Gauss-Bonnet invariant can exhibit the
phenomenon of \emph{spontaneous black hole scalarization}. These
theories admit both the classic black hole solutions predicted by
general relativity as well as novel hairy black hole solutions. The
stability of hairy black holes is strongly dependent on the precise
form of the scalar-gravity coupling.  A radial stability
investigation revealed that all scalarized black hole solutions are
unstable when the coupling between the scalar field and the
Gauss-Bonnet invariant is quadratic in the scalar, whereas stable
solutions exist for exponential couplings.  Here we elucidate this
behavior. We demonstrate that, while the quadratic term controls the
onset of the tachyonic instability that gives rise to the black hole
hair, the higher-order coupling terms control the nonlinearities
that quench that instability, and hence also control the stability
of the hairy black hole solutions.
\end{abstract}

\date{{\today}}
\maketitle
\section{Introduction}
\label{eq:intro}

A century after the inception of general relativity (GR), we have
entered the era of gravitational wave astronomy. The LIGO/Virgo
Collaboration has already observed ten binary black hole (BH) mergers
and one binary neutron star merger~\cite{LIGOScientific:2018mvr}, and
this number is only expected to grow as the sensitivity is increased
and future detectors (such as LIGO-India and KAGRA) come online. This
new observational window offers us the unprecedented opportunity to
test gravity on new distance and energy scales using some of the most
extreme objects in the
Universe~\cite{Berti:2015itd,Yunes:2016jcc,Berti:2018cxi,Berti:2018vdi,Barack:2018yly}.
Indeed, the events already observed have provided new bounds on modified
gravity theories and important consistency
tests~\cite{TheLIGOScientific:2016src,Sakstein:2017xjx,Creminelli:2017sry,
Baker:2017hug,Ezquiaga:2017ekz,Crisostomi:2017lbg,Langlois:2017dyl,Dima:2017pwp,
Abbott:2018lct}.

Astrophysical BHs in GR are simple objects characterized by two
numbers: their mass and spin. Because of their simplicity, they are
attractive probes for testing GR. Any observational test of
gravity must compare the predictions of GR with those coming from
competing theories. Without alternative predictions, one cannot
verify whether GR is correct, or even quantify the amount by which
GR is the preferred theory.
This is where the challenges typically arise: the vast majority of
well-motivated modifications of GR are subject to no-hair theorems
that preclude the existence of new BH charges.
In particular, many of these models contain an additional scalar
degree of freedom $\varphi$ \cite{Berti:2015itd,Sotiriou:2015lxa},
and the no-hair theorems preclude the existence of BHs with some
new \emph{scalar charge} $Q$~\cite{Hawking:1972qk,Mayo:1996mv,Sotiriou:2011dz,
Hui:2012qt,Sotiriou:2013qea,Sotiriou:2015pka,Herdeiro:2015waa}.
A notable exception to this rule are theories that include a
coupling of the scalar field to the Gauss-Bonnet invariant
${\cal G}=R^2-4R_{ab}R^{ab}+R_{abcd}R^{abcd}$, where $R$, $R_{ab}$ and
$R_{abcd}$ are the Ricci scalar, the Ricci tensor and the Riemann
tensor, respectively.
Such scalar-Gauss-Bonnet (sGB) couplings arise in the low-energy
effective field theory (EFT) derived from string theory, and it has
long been known that they give rise to non-Schwarzschild and non-Kerr
BHs (see e.g.~\cite{Mignemi:1992nt,Kanti:1995vq}).
A linear coupling between a scalar field and the Gauss-Bonnet
invariant features prominently in the EFT of shift-symmetric
scalars~\cite{Sotiriou:2013qea}, since the Gauss-Bonnet term is a
topological invariant and a total divergence in four
dimensions.
Again, studies have found new hairy BH solutions in these
theories~\cite{Sotiriou:2013qea,Sotiriou:2014pfa,Ayzenberg:2014aka,
Maselli:2015yva,Antoniou:2017acq,Antoniou:2017hxj}.
The fact that LIGO/Virgo has observed BH merger events consistent
with GR implies that such couplings are necessarily
small~\cite{Witek:2018dmd}.

Recently, a new possibility was pointed out and named ``spontaneous BH scalarization":
certain sGB theories admit both the BH solutions of GR and hairy BH
solutions~\cite{Doneva:2017bvd,Silva:2017uqg}.
This phenomenon can occur in theories where a scalar field is coupled
to the Gauss-Bonnet invariant and the coupling function respects
$\mathbb{Z}_2$ symmetry and vanishes for some constant $\varphi_0$.
The last condition guarantees that GR BHs are admissible
solutions. The coupling with the Gauss-Bonnet invariant acts as an
effective mass term for the scalar perturbations around these
solutions. When the BH mass lies within a certain interval, this
effective mass term is negative in parts of the BH exterior,
triggering a tachyonic instability and producing a nonzero scalar
charge.
The scalarized solutions exist in BH mass bands whose onset
coincides with the tachyonic instability, and whose termination is
due to regularity conditions on the horizon that arise from
nonlinear effects~\cite{Doneva:2017bvd,Silva:2017uqg}.
Similar scalarization phenomena have been studied also for neutron
stars~\cite{Silva:2017uqg,Doneva:2017duq}, Reissner-Nordstr\"om and
Kerr BHs~\cite{Cardoso:2013opa,Cardoso:2013fwa,Doneva:2018rou,
Herdeiro:2018wub,Brihaye:2018bgc}, and scalar-tensor gravity coupled
with Born-Infeld electrodynamics~\cite{Stefanov:2007eq,Doneva:2010ke}.

There is no \emph{a priori} guidance for the functional dependence of the
coupling function $f(\varphi)$. Reference~\cite{Silva:2017uqg} focused on
the quadratic coupling $f(\varphi)\sim \varphi^2$, as this is the
simplest case where the tachyonic instability should be present and
the leading-order term is expected to control the onset of the
instability. Reference~\cite{Doneva:2018rou} focused on the exponential
coupling $f(\varphi)\sim\exp(\beta\varphi^2)$ instead.
Recent studies suggest that scalarized BH solutions are unstable
under radial perturbations for the quadratic coupling, while
solutions within the exponential coupling model have better stability
properties~\cite{Blazquez-Salcedo:2018jnn}.

This paper is concerned with understanding the nature of the
instability of the quadratic coupling function. In particular, we will
show that the radial instability of the quadratic model is directly
linked to the fact that, in this model, the scalar field equation is
linear in the scalar. This implies that gravitational backreaction is
crucial for quenching the tachyonic instability that leads to
scalarization, and that backreaction determines the properties of the
scalarized solution in this model. Here we find that introducing
nonlinearity in the scalar provides a different quenching mechanism
for the tachyonic instability, changes the properties of the
scalarized solutions, and removes the radial instability. The simplest
setup to demonstrate these points is a theory where the quadratic
coupling is augmented by a quartic term\footnote{The $\varphi^4$
correction to the coupling function introduces nonlinearity in
$\varphi$ in the scalar's equation and is the leading order term
with this property in the exponential coupling case.  Note that
from an EFT perspective, augmenting the coupling function is not
the most natural method for stabilizing the solutions. However,
our goal here is to understand the role and properties of the
coupling function. We will discuss EFT considerations in a
forthcoming publication~\cite{Macedo}.}.

The plan of the paper is as follows.
In Sec.~\ref{sec:theory} we briefly review sGB gravity, the
necessary conditions for the existence of scalarized BH solutions, and
our chosen coupling functions.
In Sec.~\ref{sec:decoupling_limit} we study quartic sGB gravity in
the decoupling limit, compute scalar field bound states and
investigate their stability.
In Sec.~\ref{sec:radial_stability} we obtain full nonlinear BH
solutions in this theory and discuss their stability under radial
perturbations.
In Sec.~\ref{sec:conclusions} we summarize our findings.

\section{Scalar-Gauss-Bonnet gravity}
\label{sec:theory}

In sGB gravity, a real, massless scalar field is coupled to gravity
through the Gauss-Bonnet invariant ${\cal G}$.
The action of sGB gravity is
\begin{equation}
S = \frac{1}{2}
\int {\rm d}^4x
\sqrt{-g}
\left[
R - \frac{1}{2}g^{ab}\varphi_{;a}\varphi_{;b} + f(\varphi){\cal G}
\right]\,,
\end{equation}
where $\varphi$ is the scalar field and $g_{ab}$ is the spacetime metric.
We use geometrical units, such that $8\pi G=c=1$. The field equation for
the scalar field in sGB gravity is
\begin{equation}
\Box \varphi = - f_{,\varphi}(\varphi){\cal G}\,,
\label{eq:eom}
\end{equation}
while the equation for the spacetime metric is
\begin{equation}
\label{eq:k}
R_{ab}-\frac{1}{2}g_{ab}R=T_{ab}
\end{equation}
where $T_{ab}$ is the sum of the matter stress-energy tensor (which is
vanishing for BH solutions) and an effective stress-energy tensor
which depends on $f_{,\varphi}(\varphi)$~\cite{Antoniou:2017acq}.
Different choices of the function $f(\varphi)$ correspond to
different sGB gravity theories.
In particular, $f(\varphi)\sim \exp(\alpha\varphi)$ (where $\alpha$ can be
different depending on the specific stringy scenario) corresponds to
Einstein-dilaton Gauss-Bonnet (EdGB) gravity~\cite{Mignemi:1992nt,Kanti:1995vq,Maeda:2009uy,Torii:1998gm,
Pani:2009wy,Yunes:2011we,Pani:2011gy,Yagi:2011xp,Yagi:2012gp,Ayzenberg:2014aka,
Maselli:2015tta,Kleihaus:2015aje,Blazquez-Salcedo:2016enn},
which can arise in the low-energy effective actions of some string
theories~\cite{Gross:1986mw,Metsaev:1987zx};
$f(\varphi)\sim\varphi$ corresponds to shift-symmetric sGB gravity,
which is invariant under $\varphi\rightarrow\varphi+{\rm constant}$,
and is so far the only known shift-symmetric scalar-tensor theory with
second-order field equations to allow for asymptotically flat, hairy BH
solutions~\cite{Sotiriou:2013qea,Sotiriou:2014pfa,Maselli:2015yva}.

Remarkably, EdGB and shift-symmetric sGB gravity do not admit
Schwarzschild BH solutions: all static, spherically symmetric BH
solutions in this theory have nontrivial scalar field configurations.
This feature, however, is not shared by all sGB gravity theories.
As shown in~\cite{Doneva:2017bvd,Silva:2017uqg}, sGB gravity
admits the BH solutions of GR (with a constant scalar field)
if $f_{,\varphi}(\varphi_0)=0$ for some constant $\varphi_0$, and
it {\it also} admits BH solutions with nontrivial scalar field
configurations if $f_{,\varphi\varphi}{\cal G}<0$.
In these theories, the $f_{,\varphi\varphi}{\cal G}$ term in the
field equations acts as a negative mass term, triggering a tachyonic
instability\footnote{See also~\cite{Myung:2018iyq} for an alternative
interpretation in terms of the Gregory-Laflamme instability.},
which can in principle lead to the development of ``scalar hair''.
This process of spontaneous scalarization is analogous to that
studied in compact stars in scalar-tensor gravity~\cite{Damour:1993hw,Damour:1996ke},
but, \emph{crucially}, it does not rely on any coupling with matter,
and therefore it could potentially be tested through the observation
of gravitational waves from binary BH mergers.

Two examples of sGB gravity theories satisfying the conditions for
spontaneous scalarization have been studied: quadratic sGB
gravity~\cite{Silva:2017uqg}, in which $f(\varphi)\sim\varphi^2$, and
exponential sGB gravity~\cite{Doneva:2017bvd}, where
$f(\varphi)\sim1-\exp(-3\varphi^2/2)$~\footnote{Note that our notation
is consistent with that of~\cite{Silva:2017uqg} but different from
the notation of~\cite{Doneva:2017bvd}, since the scalar field
defined in~\cite{Doneva:2017bvd} differs from our scalar field by a
factor $2$.}.

In this work we will study {\it quartic sGB gravity}, with a
coupling term of the form
\begin{equation}
f(\varphi) \equiv \frac{1}{8}\bar \eta \varphi^2 + \frac{1}{16} \bar \zeta \varphi^4 \,,
\label{eq:coupling_fun}
\end{equation}
where $\bar \eta$, $\bar \zeta$ are coupling constants with dimensions of
[length]$^2$, and the numerical factors are chosen for convenience.
The scalar field is dimensionless, while ${\cal G}$ has units of
[length]$^{-4}$. When $\bar \zeta=0$, we obtain the quadratic sGB
gravity theory considered in~\cite{Silva:2017uqg}, which allows for
spontaneous scalarization when $\bar \eta>0$.
For small values of the scalar field, the exponential sGB gravity
studied in~\cite{Doneva:2017bvd}
\begin{equation}
f(\varphi) =\frac{\lambda^2}{12}
\left[1 - \exp(-3\varphi^2/2)\right]
\label{eq:coupling_exp}
\end{equation}
reduces to quartic sGB gravity with
\begin{equation}
  \bar \eta=\lambda^2\,,\quad
  \bar \zeta=-\frac{3}{2}\lambda^2\,,
  \label{eq:donez}
\end{equation}
plus ${\cal O}(\varphi^6)$ terms.

Since the field equations~\eqref{eq:k} reduce to Einstein's equations
when $f_{,\varphi}(\varphi)=0$, quartic sGB gravity admits the GR
solutions provided that
\begin{equation}
f_{,\varphi}=\frac{\varphi}{4}\left(\bar \eta+\bar \zeta\varphi^2\right)=0.
\end{equation}
Therefore, it always admits the GR solutions with $\varphi\equiv0$,
and if $\bar \eta \bar \zeta<0$ it admits
two additional, real-valued, constant scalar field solutions
$\varphi\equiv\varphi_+$ and $\varphi\equiv\varphi_-$, where
\begin{equation}
\varphi_{\pm} = \pm \sqrt{|\bar \eta/\bar \zeta|}\,.
\label{eq:exact_const_sol}
\end{equation}

\section{Decoupling limit of quartic scalar-Gauss-Bonnet gravity}
\label{sec:decoupling_limit}
Our goal in this section is to understand whether a quartic term in
the coupling function $f(\varphi)$ i.e., a nonzero value of $\bar \zeta$
can stabilize static, spherically symmetric BH solutions, which are
known to be unstable in the quadratic case.
Since quadratic sGB gravity only admits scalarized solutions if
${\bar \eta}\sim f_{,\varphi\varphi}>0$, in the following we shall assume
${\bar \eta}>0$, as in~\cite{Silva:2017uqg,Blazquez-Salcedo:2018jnn}.

To begin, we consider the {\it decoupling limit} in which we neglect
the backreaction from the metric. Thus, we study the scalar field
equation~\eqref{eq:eom}
\begin{equation}
\Box \varphi = - f_{,\varphi}(\varphi){\cal G}=-\frac{\varphi}{4}\left(\bar \eta+\bar \zeta\varphi^2\right){\cal G}\,,
\label{eq:eom2}
\end{equation}
on a {fixed} Schwarzschild background with mass $M$.
As discussed in~\cite{Doneva:2017bvd,Silva:2017uqg}, this class of
theories admits two kinds of solutions: the GR solutions, i.e.
a constant scalar field with $\varphi\equiv0,\varphi_\pm$; and the scalarized
solutions, in which the scalar field has a nontrivial configuration.

\subsection{Static bound-state solutions}\label{sec:boundstate}
As a first step, we look for static, bound-state solutions of quartic
sGB gravity in the decoupling limit. As in Ref.~\cite{Silva:2017uqg},
we consider a time-independent scalar field and we expand it in
spherical harmonics in standard Schwarzschild coordinates:
\begin{equation}
\varphi = \frac{1}{r}\sum_{\ell m} \bar \sigma_{\ell m}(r) Y_{\ell m}(\theta,\phi)\,.
\end{equation}
Due to the nonlinearity introduced by the quartic term
in Eq.~\eqref{eq:coupling_fun}, the wave equation is only separable for
spherically symmetric configurations ($\ell = 0$). Defining
$\bar \sigma=\bar \sigma_{00}$ and introducing the
dimensionless variables
\begin{align*}
\sigma &\equiv \bar \sigma / (2M),
\quad
\,\,\rho \equiv r / (2M),
\nonumber \\
\eta &\equiv \bar \eta / (2M)^2,
\quad
\zeta \equiv \bar \zeta / (2M)^2,
\end{align*}
we obtain the following nonlinear differential equation:
\begin{align}
  \sigma'' &+ \frac{1}{(\rho - 1)}
             \left[\frac{\sigma'}{\rho}
- \frac{\sigma}{\rho^2}
+ \frac{3\eta \sigma}{\rho^5}
   + \frac{3\zeta \sigma^3}{\rho^7} \right]= 0\,,
\label{eq:scalar_eom}
\end{align}
where primes denote derivatives with respect to $\rho$.

Introducing the tortoise coordinate $\rho_{\ast} = \rho + \log(\rho - 1)$,
this equation becomes a Schr\"odinger-like equation with a {\it nonlinear}
potential.
We cannot straightforwardly apply results from quantum mechanics to
determine when Eq.~\eqref{eq:scalar_eom} admits bound-state solutions,
as done e.g. in Refs.~\cite{Dotti:2004sh,Cardoso:2010rz,Cardoso:2013fwa}
using criteria derived in~\cite{Buell:1995}, and therefore we must study
Eq.~\eqref{eq:scalar_eom} numerically.
We start the integrations close to the event horizon $\rho = 1$
(typically at $\rho = 1 + 10^{-5}$) using a power series solution for
$\sigma$ valid in the near-horizon region. To leading order we have
\begin{equation}
\sigma = \sigma_{\rm h}
+ (\sigma_{\rm h} - 3\eta\sigma_{\rm h} - 3\zeta\sigma_{\rm h}^3)
(\rho - 1)\,.
\label{eq:vtheta_near_hor}
\end{equation}
We then integrate out to a large $\rho$ (typically $\approx 10^{5}$)
for given values of $(\eta,\,\zeta)$ and an arbitrary value
$\sigma_{\rm h} \equiv \sigma(1)$.
For each choice of the coupling constants $\eta$, $\zeta$, and
for each choice of $\sigma_{\rm h}$, Eq.~\eqref{eq:scalar_eom}
admits a unique solution with
$\sigma'_{\rm h}\equiv\sigma'(1)=\sigma_{\rm h}(1-3\eta-3\zeta\sigma_{\rm h}^2)$, $\varphi_{\rm h}=\sigma_{\rm h}$
and $\varphi'_{\rm h}=-3\sigma_{\rm h}(\eta+\zeta\sigma_{\rm h}^2)$.
We find that the scalar field diverges as $\rho\rightarrow\infty$ if and
only if $\varphi_{\rm h}'/\varphi_{\rm h}>0$, i.e. if $\eta+\zeta\sigma_{\rm h}^2>0$.
Since we assume that $\eta>0$, this condition is always satisfied if $\zeta>0$,
while if $\zeta<0$ the condition is satisfied for $\varphi_-<\varphi_{\rm h}<\varphi_+$
[see Eq.~\eqref{eq:exact_const_sol}].

Since we are interested in scalarized solutions with the same
asymptotic behavior as in GR, we require that the scalar field vanishes
at infinity.
We find that this condition can only be enforced for $\eta$ larger than
a threshold value $\eta_{\rm thr}=0.726$; for each $\eta>\eta_{\rm thr}$,
there is a discrete set of values of $\sigma_{\rm h}$ which correspond to the
solutions satisfying the boundary condition at infinity.
This value, i.e. $\bar\eta_{\rm thr}/M^2=4\eta_{\rm thr}=2.904$,
coincides with the first (zero-node) eigenvalue in the quadratic
theory of Ref.~\cite{Silva:2017uqg}. We focus on the nodeless
solution because previous work \cite{Blazquez-Salcedo:2018jnn}
showed that this is the only stable scalarized solution in the
exponential theory. Therefore it is natural to ask whether the
nodeless solution is stable in the (simpler) quartic theory.

\begin{figure}[t!]
\includegraphics[width=\columnwidth]{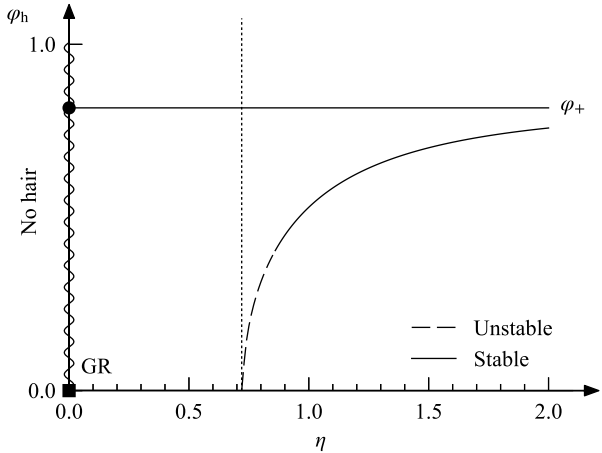}
\caption{Bound scalar field solutions (with no nodes) in the quartic
theory with $\zeta=-(3/2)\eta$.  Solutions for which the effective
potential $V_{\rm eff}$ defined in ~Eq.~\eqref{eq:veff}
is (is not) positive definite correspond to the solid (dashed) line.
The horizontal line (marked by a circle) corresponds to the constant
$\varphi_{+} = \sqrt{3/2}$ solution, whereas the vertical line
corresponds to the solutions of the quadratic sGB.
For $\eta = 0$, the theory reduces to that of scalar field minimally
coupled to gravity. No-hair theorems
force $\varphi$ to have (any) constant value.
See text for a detailed description.
}
\label{fig:solution_space}
\end{figure}

The results of the integration of Eq.~\eqref{eq:scalar_eom} are shown
in Fig.~\ref{fig:solution_space}, where we plot the scalar field at
the horizon, $\varphi_{\rm h}$, compatible with the boundary conditions,
in the case of $\zeta=-(3/2)\eta$ [corresponding to the choice in
Eq.~\eqref{eq:donez}], for the solution with no nodes.
Note that quartic sGB gravity is symmetric under $\varphi\rightarrow-\varphi$,
so we only show solutions with $\varphi>0$.
For $\eta<\eta_{\rm thr}$ the only solution is $\varphi_{\rm h}=0$, and the
scalar field is zero everywhere. A scalarized solution appears for $\eta>\eta_{\rm thr}$.
As $\eta$ increases, the value of $\varphi_{\rm h}$ for the scalarized solution also
increases, and it tends to the limit $\varphi_+$ as $\eta\rightarrow\infty$.
The same qualitative behavior occurs for different (negative) values of
$\zeta$.

In the same figure we show the corresponding curve for the quadratic
theory~\cite{Silva:2017uqg}.  In this case the scalar field equation
in the decoupling limit is linear, therefore only a discrete set of
values of $\eta$ fulfills the boundary conditions, and the zero-node
solution corresponds to $\eta=\eta_{\rm thr}$. For this value all
choices of $\varphi_{\rm h}$ are equivalent, since the solution of a linear
equation is defined modulo an overall multiplicative constant.

\subsection{Linear stability analysis in the decoupling limit}
\label{sec:stability}

Let us now analyze the stability of the static, spherically symmetric
solutions discussed in Sec.~\ref{sec:boundstate}. Let
\begin{equation}
\sigma = \sigma_0 + \delta\sigma\,,
\end{equation}
were $\sigma_0$ is a solution of Eq.~\eqref{eq:scalar_eom}
(e.g. found numerically as in Sec.~\ref{sec:boundstate}).
Substituting this perturbation in Eq.~\eqref{eq:scalar_eom} and
neglecting ${\cal O}(\delta\sigma^2)$ terms, we find the linear equation
\begin{equation}
    \frac{{\rm d}^2 \delta\sigma}{{\rm d}\rho_{\ast}^2}
- V_{\rm eff}(\rho) \,\delta\sigma = 0\,,
\label{eq:eom_perturbation}
\end{equation}
with effective potential
\begin{equation}
V_{\rm eff} \equiv
\left(1 - \frac{1}{\rho} \right)
\left[
\frac{\ell(\ell + 1)}{\rho^2} + \frac{1}{\rho^3}
- \left(\frac{3\eta}{\rho^6} + \frac{9\zeta}{\rho^8}\sigma_0^2\right)
\right]\,.
\label{eq:veff}
\end{equation}

Following~\cite{Buell:1995}, a {sufficient} (but not necessary)
condition for instability is
\begin{align}
\int_{1}^{\infty}
\frac{V_{\rm eff}}{1 - 1/\rho}\,{\rm d}\rho
&= \ell(\ell + 1) + \frac{1}{2}
- \frac{3\eta}{5} - 9\zeta \int_{1}^{\infty} \frac{\varphi_0^2}{\rho^6}\,{\rm d}\rho
   < 0\,,
\label{eq:sufficient_conditions}
\end{align}
where $\varphi_0 = \delta \sigma_0 / \rho$ and the integral must be
computed numerically.

In the quadratic sGB theory ($\zeta=0$) the solution is unstable for
$\eta>5/6\simeq0.83$.  Remarkably, in this case the effective
potential does not depend on $\varphi_0$.  Therefore, when the
Schwarzschild solution is unstable -- the instability eventually
leading to spontaneous scalarization -- the bound state solution is
also unstable.
This simple qualitative reasoning suggests that, as shown
in~\cite{Blazquez-Salcedo:2018jnn}, scalarized solutions in the
quadratic sGB theory are always unstable.

Eq.~\eqref{eq:sufficient_conditions} also shows that when $\zeta<0$
the contribution of the integral is positive, and therefore it tends
to stabilize BH solutions.
Moreover the integral term vanishes as $\varphi_0\to 0$. This suggests
that in the quartic theory, for certain values of the coupling
constants $(\eta,\,\zeta)$ the Schwarzschild solution is
unstable, while the scalarized solution is not.
This is consistent with the results found
in~\cite{Blazquez-Salcedo:2018jnn} for the exponential theory, which
is equivalent to the quartic theory with $\zeta=- (3/2) \eta$ if we
ignore terms of order ${\cal O}(\varphi^6)$.

\subsection{Numerical results}

To assess the stability of the scalarized solutions under radial
perturbations, we computed the effective potential and the integral in
Eq.~\eqref{eq:sufficient_conditions} numerically in the case
$\zeta=-(3/2)\eta$ using the bound-state solutions $\varphi_0$
corresponding to the curve in Fig.~\ref{fig:solution_space}.
We find that the condition~\eqref{eq:sufficient_conditions}
(which is just a sufficient condition for instability) is
never satisfied for the bound-state solutions.
Looking at the minimum of the effective potential, we find that it is
negative for $\eta_{\rm thr}=0.726<\eta<0.86$, while it is positive for
$\eta>0.86$.
This is an indication that the bound state solutions, at least for $\eta>0.86$,
are linearly stable under radial perturbations.
In Fig.~\ref{fig:solution_space} we mark solutions for which $V_{\rm eff}$ is
not positive-definite by a dashed line, and those for which $V_{\rm eff}$
is positive everywhere by a solid line.
In the GR limit ($\eta=0$) the theory reduces to a scalar field
minimally coupled to gravity, and no-hair theorems~\cite{Hawking:1972qk,
Mayo:1996mv,Sotiriou:2011dz} require $\varphi$ to be (any) constant,
as indicated by the wiggly line.

These results are in qualitative agreement with those
for the exponential theory, as can be seen from a comparison with
Fig.~2 of~\cite{Blazquez-Salcedo:2018jnn}.
A scalarized solution exists when $1/(2\sqrt{\eta})=M/\bar\eta^{1/2}$
(which is the same as $M/\lambda$ in their notation) is smaller than
$0.59$, but for $0.54< M/\bar\eta^{1/2}<0.59$ the effective potential
is not positive definite.

We now turn to a study of the perturbations of the fully coupled field
equations [Eqs.~\eqref{eq:eom} and \eqref{eq:k}].

\section{Nonlinear black hole solutions and their radial (in)stability}
\label{sec:radial_stability}

In GR, radial perturbations describe nonradiative fields. The
perturbation equations can be solved analytically and correspond to a
change in mass of the Schwarzschild BH solution, as expected from
Birkhoff's theorem~\cite{Zerilli:1971wd}.
In modified theories of gravity
radial perturbations can be radiative, with important consequences for
the stability of the spacetime.

As mentioned in the introduction, Ref.~\cite{Blazquez-Salcedo:2018jnn}
studied radial perturbations of a static, spherically symmetric BH for
a generic coupling function $f(\varphi)$ in sGB gravity.
We mostly follow their treatment. For brevity, we will only outline
the procedure to obtain the perturbation equations and our numerical
calculation of the radial oscillation modes.

The spherically symmetric, radially perturbed spacetime up to first
order in the perturbations has line element
\begin{align}
    {\rm d}s^2=&-\exp [2\Phi(r)+\varepsilon {F_t}(t,r)]{\rm d}t^2\nonumber\\
               &+\exp [2\Lambda(r)+\varepsilon {F_r}(t,r)]{\rm d}r^2+r^2{\rm d}\Omega^2\,,
\label{eq:metric_p}
\end{align}
where $\varepsilon$ is a small bookkeeping parameter
and ${\rm d} \Omega^2 = {\rm d}\theta^2 + \sin^2\theta\, {\rm d} \phi^2$
is the line element of the unit-sphere.
To the same order, the scalar field is given by
\begin{equation}
\varphi=\varphi_0(r)+\varepsilon \frac{\varphi_1(t,r)}{r}\,.
\label{eq:scalar_f}
\end{equation}
By inserting Eqs.~\eqref{eq:metric_p} and \eqref{eq:scalar_f} into the
field equations~\eqref{eq:eom}-\eqref{eq:k} and expanding in powers of
$\varepsilon$ we get equations for the background metric functions
$(\Phi,\,\Lambda)$ at zeroth order in $\varepsilon$, and for the
radial perturbations $(F_t,\,F_r,\,\varphi_1)$ at first order in
$\varepsilon$. Let us first discuss the background equations and
boundary conditions.

\subsection{Black hole scalarization in quartic sGB gravity}
\label{sec:nonlinear_bhs_quartic}

The zeroth-order equations for $(\Phi,\,\Lambda,\,\varphi_0)$ can be
cast as a coupled system of two first-order equations for $\Phi$ and
$\Lambda$ and a second-order equation for
$\varphi_0$~\cite{Kanti:1995vq,Doneva:2017bvd,Blazquez-Salcedo:2018jnn}.
BH solutions are obtained by imposing that the metric functions
$\exp(2\Phi)$ and $\exp(-2\Lambda)$ vanish and that the scalar field
$\varphi_0$ be regular at the horizon:
\begin{align}
\exp(2\Phi) &\sim (r-r_{\rm h})+{\cal O}[(r-r_{\rm h})^2]\,,\label{eq:bc_hor1}\\
\exp(-2\Lambda) &\sim(r-r_{\rm h})+{\cal O}[(r-r_{\rm h})^2]\,,\label{eq:bc_hor2}\\
\varphi_0 &\sim\varphi_{0,{\rm h}}+\varphi_{0,{\rm h}}'(r-r_{\rm h})+{\cal O}[(r-r_{\rm h})^2]\,,\label{eq:bc_hor3}
\end{align}
where $r_{\rm h}$ is the horizon radius, while $\varphi_{0,{\rm h}}$
and $\varphi_{0,{\rm h}}'$ denote the scalar field and its first
derivative at the horizon.
Using a near-horizon expansion of the field equations, we find that BH
solutions correspond to the condition
(cf.~\cite{Antoniou:2017acq,Silva:2017uqg,Doneva:2017bvd} or more
details)
\begin{align}
\varphi_{0,{\rm h}}'=-\frac{r_{\rm h}}{\varphi_{0,{\rm
  h}}\left(\eta+\zeta \varphi_{0,{\rm h}}^2\right)}
\left[1-\sqrt{1-\frac{6\varphi_{0,{\rm h}}^2}{r_{\rm h}^4}\left(\eta+\zeta \varphi_{0,{\rm h}}^2\right)^2}\right]\,.
\nonumber \\
\label{eq:der_field}
\end{align}
When Eq.~\eqref{eq:der_field} is not satisfied, $\varphi''_{0}$ diverges at the horizon.
By requiring that the first derivative of the scalar field at the
horizon be real and using Eq.~\eqref{eq:der_field} we find a condition
for the existence of the solutions:
\begin{equation}
6\varphi_{0,{\rm h}}^2\left(\eta+\zeta \varphi_{0,{\rm h}}^2\right)^2<r_{\rm h}^4\,.
\label{eq:real_cond}
\end{equation}

At large distances, an expansion of the background equations in powers
of $r^{-1}$ leads to
\begin{align}
\exp(2\Phi) &\sim 1-\frac{2M}{r}+{\cal O}(r^{-2})\,,\label{eq:bc_inf1}\\
\exp(-2\Lambda) &\sim 1-\frac{2M}{r}+{\cal O}(r^{-2})\,,\label{eq:bc_inf2}\\
\varphi_0 &\sim \varphi_{0,\infty}+\frac{Q}{r}+{\cal O}(r^{-2})\,,\label{eq:bc_inf3}
\end{align}
where $M$ is the total (ADM) mass and $Q$ is the scalar charge. Since
we are not interested in cosmological effects and we require that the
hairy solution has the same asymptotic properties as the GR solution,
we will also impose $\varphi_{0,\infty}=0$.

To obtain scalarized BH solutions we proceed as follows.  We integrate
the differential equations for the background from the horizon
outwards imposing the boundary
conditions~\eqref{eq:bc_hor1}--\eqref{eq:bc_hor3} with the
constraint~\eqref{eq:real_cond}, using a guess value for the scalar
field at the horizon $\varphi_{0,{\rm h}}$.
Numerical solutions in the far-horizon region ($r\gg r_{\rm h}$) are
then compared with the boundary
conditions~\eqref{eq:bc_inf1}--\eqref{eq:bc_inf2}.
Not all sets of $(\varphi_{0,\rm{h}},\eta,\zeta)$ allow for BH
solutions satisfying both boundary conditions. This generates a
boundary value problem that can be solved by a shooting method. In
practice, we fix the values of $(\eta,\zeta)$, and we find the values
of $\varphi_{0,{\rm h}}$ by shooting and requiring that the scalar
field vanishes in the far region.

\subsection{Radial perturbations of scalarized black holes}
\label{sec:radial_pert}

Let us now consider the radial stability of the BH solutions found in
the preceding section.  By manipulating the first-order equations we
can show that the functions $(F_t,\,F_r,\,\varphi_1)$ are not
independent: $F_t$ and $F_r$ can be written in terms of $\varphi_1$,
where $\varphi_1$ obeys the differential
equation~\cite{Torii:1998gm,Blazquez-Salcedo:2018jnn}
\begin{equation}
h(r)\frac{\partial^2\varphi_1}{\partial t^2}-\frac{\partial^2\varphi_1}{\partial r^2}+k(r)\frac{\partial\varphi_1}{\partial r}+p(r)\varphi_1=0\,.
\label{eq:scalar_eq1}
\end{equation}
Here $h$, $k$ and $p$ are functions of $r$ that depend on the
background metric functions: cf. Eq. (14) of~\cite{Blazquez-Salcedo:2018jnn}.

By a suitable redefinition of the functions $(h,k,p)$ and using
a harmonic-time decomposition
$\varphi_1(t,r)=\varphi_1(r) e^{-i\omega t}$,
we can write the above equation in a Schr\"odinger-like form.
This is useful for analyzing the effective potential felt by the
perturbations~\cite{Blazquez-Salcedo:2018jnn}.
However here we will deal directly with the differential equation in
the form~\eqref{eq:scalar_eq1}, mainly because it is simpler to solve
it numerically.
We introduce a compactified dimensionless coordinate
\begin{equation}
x \equiv 1-r_{\rm h}/r\,,
\end{equation}
such that the horizon and spatial infinity are mapped to
$x=0$ and $x=1$, respectively.
To integrate Eq.~\eqref{eq:scalar_eq1} numerically we impose the
standard boundary conditions at the horizon and at spatial infinity:
\begin{equation}
\varphi_1(x)=\left\{
\begin{array}{ll}
e^{i\omega r_*} & r_*\to\infty ~~(x \to 1)\\
e^{-i\omega r_*} & r_*\to-\infty ~~ (x \to 0)
\end{array}\right.
\label{eq:bc_pert},
\end{equation}
where $r_*$ is the tortoise coordinate.

The differential equation~\eqref{eq:scalar_eq1} together with
boundary conditions~\eqref{eq:bc_pert} yields a boundary-value
problem for the complex eigenvalue $\omega=\omega_R+i\omega_I$.
Stable modes have $\omega_I<0$, while unstable modes have $\omega_I>0$.
Therefore, to study the radial stability of the solutions we can
search for purely imaginary
modes with $\omega_I>0$.
To obtain these modes, we use again a shooting method. We perform two
integrations starting at $x=0$ and at $x=1$.
At each boundary we impose that the scalar field is zero and that its
first derivative is constant. We can fix the scalar field amplitude to
unity because Eq.~\eqref{eq:scalar_eq1} is linear.  The integration of
Eq.~\eqref{eq:scalar_eq1} from the horizon yields a first solution
$\varphi_1^{(-)}$, and the integration from infinity yields a second
solution $\varphi_1^{(+)}$. We match the two solutions at an
intermediate point $x_m$.
The eigenvalue $\omega=i\omega_I$ corresponds to the
frequency at which the Wronskian
\begin{equation}
    W=\left[\varphi_1^{(-)} \frac{{\rm d}\varphi_1^{(+)}}{{\rm d}x}-\varphi_1^{(+)} \frac{{\rm d}\varphi_1^{(-)}}{{\rm d}x}\right]_{x=x_m}
\label{eq:wroskian}
\end{equation}
vanishes. We checked that the modes are stable under variations of the
numerically chosen values of the near-horizon radius, of the large
radius representing spatial infinity, and of the matching point $x_m$.
Additionally, we checked that our results reproduce those for the
exponential and quadratic couplings presented in
Ref.~\cite{Blazquez-Salcedo:2018jnn}.

\subsection{Numerical results}
\label{sec:stability_numerical}

\begin{figure}[t!]
\includegraphics[width=\columnwidth]{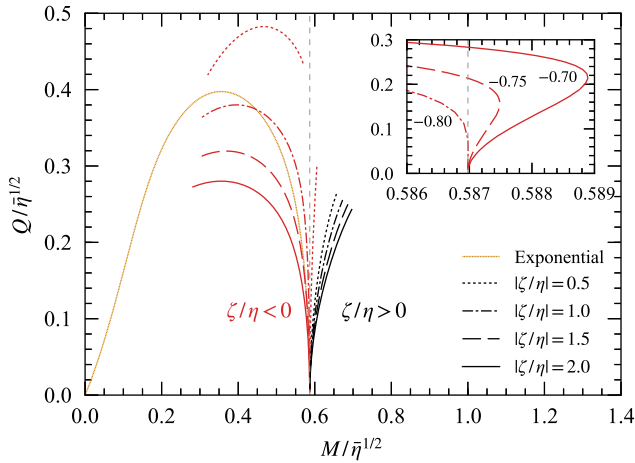}
\caption{$Q$--$M$ diagram for scalarized solutions in quartic sGB
gravity with $n=0$, considering different values for
$\zeta/\eta$. For comparison, we also show the solutions for
exponential sGB gravity [cf.~Eq.~\eqref{eq:coupling_exp}]: for small
$Q / \bar{\eta}^{1/2}$ (i.e. small scalar field amplitudes), the
curve overlaps with the case $|\zeta / \eta| = 1.5$, as it
should. The vertical line represents the scalarization threshold
$\eta=\eta_{\rm thr}$. Solutions to the left (right) of the vertical
dotted line are stable (unstable).  The inset shows additional
illustrative curves showing the behavior near the scalarization
threshold $\eta_{\rm thr}$.  Curves with
$\zeta / \eta \lesssim -0.8$ are always to the left of the
scalarization threshold.  }
\label{fig:bhs_sols}
\end{figure}

In Fig.~\ref{fig:bhs_sols} we plot the scalar charge-mass ($Q$--$M$)
diagram corresponding to nodeless ($n=0$) scalarized BH solutions in
the quartic theory for fixed values of
$|\zeta/\eta|=0,\,0.5,\,1,\,1.5$ and $2.0$. For completeness we also
show the corresponding diagram for the exponential coupling from
Ref.~\cite{Doneva:2017bvd}. The vertical line represents the threshold $\eta=\eta_{\rm thr}$.

Numerically we found that solutions with $\zeta/\eta \lesssim -0.8$
bend to the left in the $Q-M$ diagram (and so do solutions
corresponding to the exponential coupling), while solutions with
$\zeta/\eta$ larger than this critical value bend to the right.
Solutions illustrating this behavior near this critical value are
shown in the inset of Fig.~\ref{fig:bhs_sols}.
This different behavior corresponds to different radial stability
properties.
When $\zeta/\eta=-0.5$ there is a gap in the
parameter space in which BH solutions do not exist: the first
derivative of the scalar field at the horizon is complex in this
region.
This happens because Eq.~\eqref{eq:real_cond} cannot be satisfied
for parameter choices that lie in the gap.
We verified that the gap exists in the parameter range
$-0.64<\zeta/\eta<0$, and it is indeed related to the polynomial form
of the existence condition given by Eq.~\eqref{eq:der_field}.

The two different branches of solutions that exist (e.g.) when
$\zeta/\eta=-0.5$ present different stability properties, one being
stable and the other unstable.
We can understand this behavior qualitatively using intuition built
from
Sec.~\ref{sec:decoupling_limit}. From~\eqref{eq:sufficient_conditions},
we know that not only the magnitude of the coupling parameter $\zeta$,
\textit{but also the amplitude of the background solution} $\varphi_0$
contributes to quenching the instability.
In the unstable segment of the $\zeta / \eta = -0.5$ solutions, the
scalar field has small amplitude (i.e., small scalar charge) and
therefore it cannot quench the instability. This is not the case for
more charged solutions, which are stable.

\begin{figure}[t]
\includegraphics[width=\columnwidth]{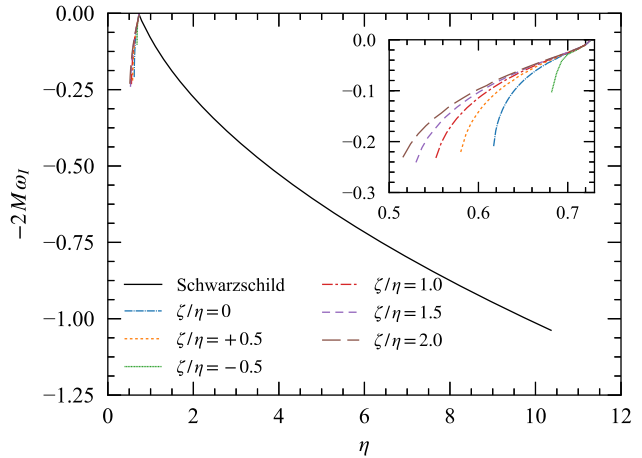}%
\caption{Eigenfrequencies of the unstable modes $\omega_I$ as
function of the coupling $\eta$ for different $\zeta/\eta$ and
for the Schwarzschild BH in sGB gravity.
The Schwarzschild BH is \textit{unstable} for $\eta\gtrsim0.726$,
as indicated by the linear analysis outlined in Sec.~\ref{sec:decoupling_limit}.
For $\eta\lesssim 0.726$ we can see the instability time scale for
the nodeless solutions for the cases $\zeta/\eta>-0.5$.
The inset zooms-in into these frequencies.
}
\label{fig:unstable_modes}
\end{figure}

To analyze the stability, in Fig.~\ref{fig:unstable_modes} we plot the
unstable mode frequencies for the same set of theories with $(\eta,\zeta)$
shown in Fig.~\ref{fig:bhs_sols}, and also for Schwarzschild BH solutions in
the same theory. Note that for Schwarzschild BH solutions
$\varphi_0=0$, and therefore the frequency does not depend on
$\zeta$. Hence, all unstable scalarized BHs branch out of the same
threshold value of $\eta\approx 0.726$, in agreement with the value
obtained in Sec.~\ref{sec:decoupling_limit}.  All BH solutions in
quartic theories with $\zeta/\eta>0$ shown in
Fig.~\ref{fig:unstable_modes} are unstable to radial perturbations,
again in agreement with the analysis from the decoupling limit, and
the instability time scale $\tau=|\omega_I^{-1}|$ of these modes
decreases as $\zeta/\eta$ increases (as long as $\zeta/\eta>0$).

As noted when discussing Fig.~\ref{fig:bhs_sols}, there is a gap
in the parameter space of BH solutions with $\zeta/\eta=-0.5$. The two
branches have different behavior in the $Q$--$M$ plane: the branch
with small values of $Q/\bar{\eta}^{1/2}$ is more similar to solutions
with $\zeta/\eta>-0.5$, and the branch with large values of
$Q/\bar{\eta}^{1/2}$ is more similar to solutions with
$\zeta/\eta<-0.5$.  We performed a radial stability analysis searching
for unstable modes in the two branches. We found no unstable modes in
the large $Q/\bar{\eta}^{1/2}$  branch, but we found unstable
modes in the small $Q/\bar{\eta}^{1/2}$ branch, and these are
the $\zeta/\eta=-0.5$ modes shown in Fig.~\ref{fig:unstable_modes}.

The case $\zeta/\eta=-0.5$, presenting stable and unstable solutions,
is similar to the ones shown in the inset of Fig.~\ref{fig:bhs_sols}.
Note that the unstable modes for $\zeta/\eta=-0.5$ have small $M\omega_I$ (see inset of
Fig.~\ref{fig:unstable_modes}).
For the solutions with values of $\zeta/\eta$ presented in the inset in
Fig.~\ref{fig:bhs_sols} the mode frequency is even smaller, being
challenging to find numerically. Our numerical findings suggest that in
quartic sGB with $\zeta/\eta<-0.8$ scalarized BHs with $n=0$ are stable.

\section{Conclusions}
\label{sec:conclusions}

In this work we have investigated the radial stability of scalarized
BH solutions in sGB gravity.  Motivated by the radial instability of
quadratic sGB solutions found in~\cite{Blazquez-Salcedo:2018jnn}, we
have shown that adding higher-order (quartic) corrections to the
original quadratic sGB model of~\cite{Silva:2017uqg} can stabilize the
solutions.

Our analysis provided a clear picture for the physical interpretation
of this results. At the linearized level, scalarization manifests as a
tachyonic instability that triggers the growth of the scalar
field. For the end-point of the instability to be a hairy solution,
the tachyonic instability needs to be quenched by some nonlinear
effects. In quadratic sGB gravity however, the field equation for the
scalar field is linear in the scalar
and hence the only quenching mechanism would be backreaction. This is
nicely demonstrated by our decoupling limit analysis, where
backreaction is entirely ignored and the tachyonic instability is
always present. Nonetheless, within the same approximation, a
higher-order coupling introduces nonlinearity in the scalar and
provides strong quenching for the tachyonic instability.  This
highlights that the very existence of hairy solutions found
in~\cite{Silva:2017uqg} for the purely quadratic model
relies on backreaction effects, and this is what renders them rather
special. This seems to be reflected on their radial stability
properties.

Indeed, a $\varphi^4$-term turns out to stabilize scalarized BH
solutions.  We computed unstable radial modes and we found none when
the coupling parameters satisfy $\zeta / \eta < -0.8$.  This suggests
that scalarized BH solutions are stable in this region of the theory's
parameter space.

More generally, our results clearly demonstrate that the quadratic
coupling between the scalar and the Gauss--Bonnet invariant controls
the onset of the scalarization, whereas the higher-order corrections
(in the scalar) control the end-point of the tachyonic instability
that triggers scalarization, and hence they are crucial for the
properties of the hairy black holes solutions.

\emph{Note Added} -- While this work was being completed, a preprint with
similar conclusions appeared as an e-print~\cite{Minamitsuji:2018xde}.
Where our works overlap, our conclusions agree with theirs.

\section*{Acknowledgments}
This work was supported by the H2020-MSCA-RISE-2015 Grant No.
StronGrHEP-690904 and by the COST action  CA16104 ``GWverse''.
H.O.S was supported by NASA Grants No.~NNX16AB98G and No.~80NSSC17M0041.
T.P.S. acknowledges partial support from the STFC Consolidated Grant
No. ST/P000703/1.
J.S. was supported by funds provided to the Center for
Particle Cosmology by the University of Pennsylvania.
E.B. is supported by NSF Grant No. PHY-1841464, NSF Grant
No. AST-1841358, NSF-XSEDE Grant No. PHY-090003, and NASA ATP Grant
No. 17-ATP17-0225.
C.F.B.M. would like to thank the Johns Hopkins University for kind
hospitality during the preparation of this work and the American
Physical Society which funded the visit through the International
Research Travel Award Program.

\bibliographystyle{apsrev4-1}
\bibliography{biblio}

\end{document}